\documentclass[12pt,letterpaper]{article}
\usepackage[a4paper, total={7in, 10in}]{geometry}

\usepackage{graphicx}
\usepackage{helvet}
\usepackage{authblk}
\usepackage{hyperref}
\usepackage{amsmath} 
\usepackage{amssymb} 
\usepackage{orcidlink} 
\usepackage[super,comma,sort&compress]{natbib}
\usepackage{hyperref}
\usepackage{url}
\usepackage[right]{lineno} \linenumbers
\usepackage[title]{appendix}%
\usepackage{booktabs}   
\usepackage{longtable}  
\usepackage{multirow}   
\usepackage{array}      
\usepackage[utf8]{inputenc} 
\usepackage[T1]{fontenc}    
\usepackage{url}            
\usepackage{booktabs}       
\usepackage{amsfonts}       
\usepackage{nicefrac}       
\usepackage{microtype}      
\usepackage{xcolor}         
\usepackage{multirow}
\usepackage{threeparttable}
\usepackage{amsmath}
\usepackage{subcaption}
\usepackage{makecell}
\usepackage{placeins}
\usepackage[utf8]{inputenc} 
\usepackage{booktabs}      
\usepackage{siunitx}       
\usepackage[font=small]{caption}       
\usepackage{tabularx}
\usepackage{listings}    
\usepackage{pifont}
\usepackage{array}
\usepackage{makecell}
\theadset{\bfseries}

\makeatletter
\renewcommand{\maketitle}{\bgroup\setlength{\parindent}{0pt}
\begin{flushleft}
  \textbf{\@title}
  
  \@author
\end{flushleft}\egroup}
\makeatother

\title{Opportunistic Screening of Wolff-Parkinson-White Syndrome
using Single-Lead AI-ECG Mobile System: A Real-World Study
of over 3.5 million ECG Recordings in China}
\date{}

\author[1,$\dagger$]{Shun Huang}
\author[2,$\dagger$,\orcidlink{0000-0002-4041-3083}]{Deyun Zhang}
\author[3]{Sumei Fan}
\author[1,5]{Gongzheng Tang}
\author[2]{Shijia Geng}
\author[1]{Yujie Xiao}
\author[4]{Xingliang Wu}
\author[1,5]{Mingke Yan}
\author[1,6]{Haoyu Wang}
\author[7]{Rui Zhang}
\author[8]{Zhaoji Fu}
\author[1,5,*,\orcidlink{0000-0001-7521-5127}]{Shenda Hong}

\affil[1]{National Institute of Health Data Science, Peking University, Beijing, China}
\affil[2]{Heart Voice Medical Technology, Hefei, China}
\affil[3]{College of Integrative Chinese and Western Medicine, Anhui University of Chinese
Medicine, Hefei, China}
\affil[4]{Tianjin Key Laboratory of Ionic-Molecular Function of Cardiovascular Disease, Department of Cardiology, Tianjin Institute of Cardiology, the Second Hospital of Tianjin Medical University, Tianjin 300211, China}
\affil[5]{Institute of Medical Technology, Peking University Health Science Center, Beijing, China }
\affil[6]{University of Chinese Academy of Sciences, China}
\affil[7]{Division of Life Science and Medicine, University of Science and Technology of China, Hefei, China}
\affil[8]{School of Management, University of Science and Technology of China, Hefei, China}

\affil[*]{Correspondence: hongshenda@pku.edu.cn}

\begin{document}

\maketitle
\insert\footins{\noindent\footnotesize $^\dagger$These authors contributed equally to this work.}
\section*{ABSTRACT}
Wolff–Parkinson–White (WPW) syndrome is a congenital cardiac conduction abnormality with a low prevalence but a significant risk of sudden cardiac death. Its early identification in the general population has long been hindered by screening costs, the scarcity of professional interpretation resources, and limited medical accessibility. Artificial Intelligence-enabled portable electrocardiography (AI-ECG) offers new possibilities for large-scale opportunistic screening; however, systematic evidence regarding its overall performance and public health value in real-world settings remains insufficient. This retrospective real-world study analyzed 3,566,626 single-lead ECG records from 87,836 individuals between 2019 and 2025 to systematically evaluate an integrated AI-ECG mobile screening system—comprising "portable devices, AI primary screening, and cardiologist review"—for WPW screening. The results demonstrate that the AI model achieved a discrimination of AUC 0.6676 and a specificity of 95.92\% in complex real-world signal environments. Despite an overall bias in predictive probabilities due to the ultra-low prevalence context, the model exhibited highly concentrated high-confidence scores for true positive individuals in high-risk groups, reflecting a stable and effective risk stratification capability. The risk of detecting WPW in AI-positive records was 86.2-fold higher than in AI-negative records.  By implementing a human-AI collaborative workflow, the volume of ECGs requiring manual review was reduced by approximately 99.5\% compared to universal screening. In an ideal collaborative scenario, an average of only 18 ECGs needed review to confirm one case of WPW, representing a more than 60-fold increase in screening efficiency. Compared to traditional hospital-based 12-lead ECGs and electrophysiological studies (EPS), this AI-ECG mobile screening system significantly reduced both time and direct medical costs. Based on large-scale real-world data, this study suggests that a risk-stratification-based human-AI collaborative screening system can reshape screening models for low-prevalence, high-risk diseases like WPW, providing a promising new paradigm for the early public health detection of rare arrhythmias.


\section*{INTRODUCTION}
Wolff–Parkinson–White (WPW) syndrome is a congenital cardiac conduction abnormality characterized by the presence of an accessory pathway (the bundle of Kent) between the atria and ventricles. This pathway bypasses the atrioventricular node, leading to pre-excitation of the ventricles\cite{R1}. While many WPW patients present typical features on an electrocardiogram (ECG), a significant portion remains asymptomatic, including 65\% of adolescents and 40\% of adults over age 30\cite{R2,R3}. However, the most severe complication of WPW syndrome is sudden cardiac death (SCD), which is often the first clinical manifestation in otherwise healthy young individuals\cite{R4}. Although the lifetime risk of SCD is 3\%–4\%\cite{R5,R6}, its potential threat to public health is profound, making the early identification and intervention of high-risk individuals a critical strategy for preventing malignant SCD events.

Nevertheless, large-scale screening for WPW faces significant challenges in clinical practice. Although invasive electrophysiological study (EPS) is the gold standard for diagnosis\cite{R7,R8, R9, R10}, its invasiveness makes it unsuitable for population-wide screening. Non-invasive methods, such as 12-lead ECG or Holter monitoring, can assist in screening but depend heavily on specialized equipment and interpretation by cardiologists\cite{R9, R10, R11, R12}, limiting their accessibility in the general population. Given the low overall prevalence of WPW (approximately 1–3 per 1,000)\cite{R9, R10, R13} and the relatively limited incidence of SCD, traditional population-wide screening strategies are not cost-effective. Furthermore, existing epidemiological evidence is largely derived from clinical or hospitalized populations, suggesting that the true burden of WPW in the general population may be chronically underestimated\cite{R14}.

In recent years, artificial intelligence has been widely applied in clinical research\cite{Zhao2024, Zhang2023, Liu2021} and has also created new opportunities for AI-enhanced portable devices in cardiac monitoring, extending application scenarios from clinical environments to daily life. The integration of AI algorithms with portable devices has made electrocardiogram (ECG) acquisition and analysis more convenient and cost-effective\cite{R15}. Meanwhile, deep learning models based on single-lead ECGs have demonstrated diagnostic performance comparable to that of cardiologists\cite{R16, R17, R18, R19}. In the context of WPW screening, this technological paradigm substantially increases the likelihood of capturing characteristic ECG patterns in real-world settings, such as PR interval shortening, delta waves, and QRS complex widening\cite{R20,R21}. This approach not only improves the accessibility of healthcare services but also offers a potential pathway to advancing health equity in resource-limited regions through home-based health management.

While large-scale real-world studies have demonstrated the feasibility of this approach for atrial fibrillation screening\cite{R22, R23, R24, R25, R26}, research for WPW syndrome remains scarce. Moreover, existing studies focus primarily on accuracy in laboratory settings rather than comprehensively evaluating the balance between clinical and cost-effectiveness in unconstrained real-world settings.There is a critical need to determine whether these capabilities can reduce unnecessary healthcare utilization, enhance patient convenience, and compensate for limitations in conventional diagnostic pathways, thereby genuinely improving patient care in real-world practice.

In our previous work, we developed WenXinWuYang, a single-lead AI-ECG mobile screening system approved by the National Medical Products Administration (NMPA, formerly CFDA)\cite{R27}. This system integrates portable ECG device, AI primary screening models, and a cardiologist review workflow, and has been successfully applied to atrial fibrillation screening. Additionally, we have developed various ECG foundation models. Among them, CardioLearn provides cloud-based deep learning services for cardiovascular disease detection\cite{R28}, while ECGFounder was trained on more than ten million ECG recordings as a large-scale foundational model\cite{R16}. The capabilities of these models have been deployed across the associated devices.

Building upon this foundation, the present study utilizes real-world data from 87,836 individuals and over 3.5 million single-lead ECGs to systematically evaluate an integrated mobile screening system comprising portable device, AI-based primary screening, and cardiologist review. Beyond traditional algorithmic performance, our evaluation treats the entire workflow as the core of clinical effectiveness, systematically assessing its collaborative efficiency and validity. Through this study, we aim to provide robust evidence supporting the public health value of this new screening paradigm, particularly for the early detection and prevention of WPW syndrome and other cardiovascular diseases.

\section*{Methods}
The primary unit of analysis in this study is the individual ECG record. This approach was chosen to evaluate the diagnostic performance and operational efficiency of the AI-ECG screening system on a per-screening basis. Each ECG record represents an independent screening event, encompassing signal acquisition, AI-based interpretation, and result delivery. Therefore, analyzing data at the ECG level provides the most direct and accurate assessment of system functionality.

\subsection*{AI-ECG System Overview}
The AI-ECG mobile screening system evaluated in this study is a comprehensive system integrating portable ECG device, AI primary screening models, and a cardiologist review workflow (Figure 1). The process begins with the user independently recording a 30-second to 5-minute ECG signal using a portable single-lead device. These signals are uploaded via a smartphone application to a cloud server, where deployed AI models perform real-time analysis and WPW risk classification. Results are provided to the user in real-time, and the system allows users to request a review by a cardiologist through the app, establishing a closed-loop service from AI-driven opportunistic screening to expert-confirmed diagnosis.

\begin{figure}[htbp] 
    \centering
    \includegraphics[width=0.8\textwidth]{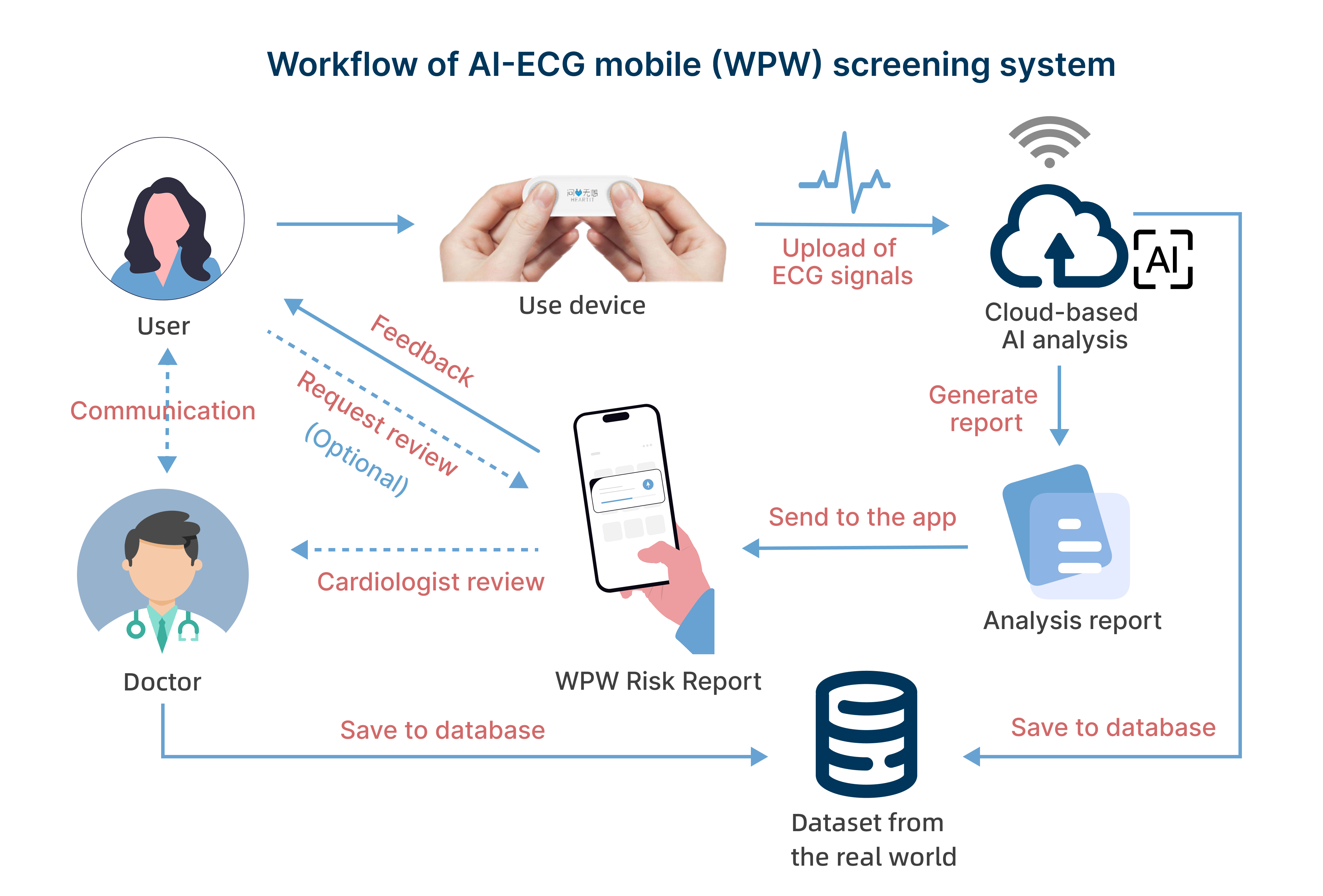} 
    \caption{Schematic overview of the AI-ECG mobile screening system workflow.}
    \label{system}
\end{figure}

\subsection*{Study Design and Data Sources}
This study was designed as a retrospective real-world study based on a large-scale, multi-regional dataset. Data were sourced from the HeartVoice-ECG database, including 3,566,626 single-lead ECG records uploaded by 87,836 individuals across China between August 2019 and May 2025. The dataset was pre-processed to exclude low-quality and invalid records. This study was approved by the Peking University Biomedical Ethics Committee (Approval No. IRB00001052-23071), and the requirement for informed consent was waived. 

A multi-layer data stratification strategy was employed (Figure 2). In the first layer, the dataset was stratified based on AI model output into two primary cohorts: "AI-positive for WPW" and "AI-negative for WPW." The second layer these cohorts were further subdivided based on whether participants actively requested cardiologist review. In the third layer, AI-negative records undergoing user-initiated review were categorized into sinus normal and other cardiac conditions.

\begin{figure}[htbp] 
    \centering
    \includegraphics[width=0.8\textwidth]{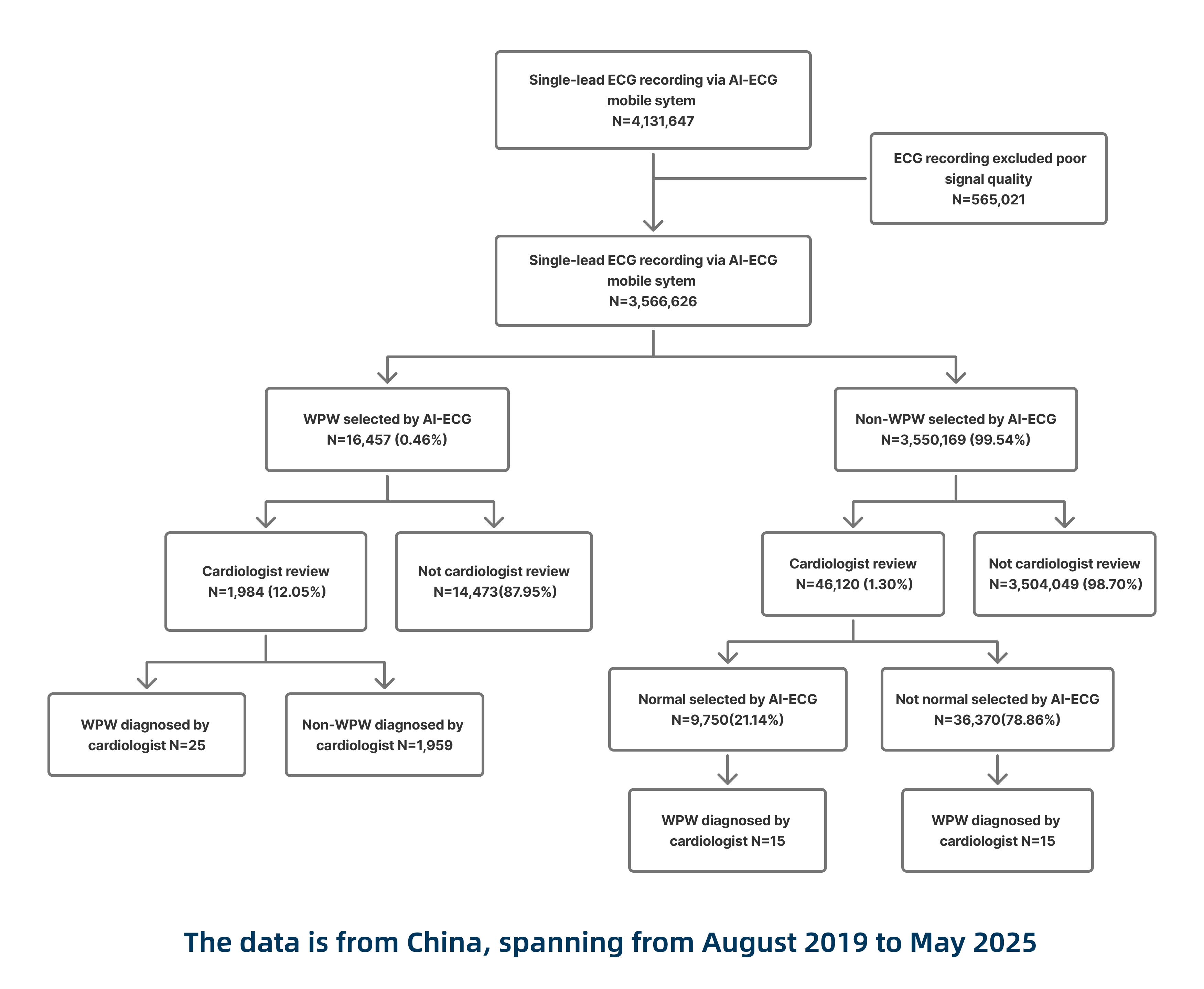} 
    \caption{Study flowchart and multi-layer data stratification strategy.}
    \label{workflow}
\end{figure}

\subsection*{Evaluation Dimensions and Outcomes}
This study adopted a multidimensional integrated evaluation framework to comprehensively assess the real-world performance of an AI-ECG mobile screening system for opportunistic WPW screening across four domains: technical performance, user behavior, clinical value, and socioeconomic impact, with particular emphasis on the collaborative workflow between AI-based primary screening and cardiologist review.

\subsubsection*{1.AI Model Performance}
We calculated basic performance metrics including AUC, sensitivity, and specificity. We also analyzed the impact of comorbidities by comparing the sensitivity and specificity between WPW and non-WPW cases with and without concurrent other cardiac conditions. 

Calibration curves and Brier scores were used to assess the consistency between predicted probabilities and actual prevalence. Given the extremely low prevalence of WPW in real-world screening settings and the primary role of the model in risk stratification, conventional global calibration may be affected by severe class imbalance. We introduced a score stability analysis for true positives to characterize confidence stratification.

\subsubsection*{2.User-initiated cardiologist review behavior}
To evaluate the real-world feasibility of AI–cardiologist collaboration, we tracked whether users requested expert review after receiving AI classification. The distribution of ECG outcomes—WPW, other cardiac conditions, or sinus normal—was calculated for reviewed and non-reviewed cases. To correct for selection bias in positive predictive value (PPV) calculations, 500 AI-positive records without user-initiated review were randomly sampled and independently annotated by three senior cardiologists. Only ECGs unanimously identified as typical WPW were confirmed as positive.

\subsubsection*{3.Value of Human-AI Collaboration}
Relative risk (RR)\cite{R29} was used to quantify the association between AI screening results (exposure: AI-positive vs AI-negative) and final cardiologist-confirmed WPW diagnosis (event). RR reflects the increased likelihood of a confirmed WPW diagnosis in AI-positive cases. 
\begin{equation}
RR = \frac{\text{Exposed risk}}{\text{Unexposed risk}} = \frac{P(\text{Event} \mid \text{Exposed})}{P(\text{Event} \mid \text{Unexposed})}
\end{equation}
To assess operational impact, the number of ECGs requiring cardiologist review was compared between baseline (non-AI) and AI-assisted scenarios, quantifying the reduction in manual workload.

\subsubsection*{4.System efficiency and economic burden}
Efficiency was evaluated using the number needed to review (NNR), defined as the average number of ECGs requiring cardiologist review to confirm one WPW case. Five screening strategies were compared: (1) Reviewing only AI-positive ECGs; (2) Reviewing user-initiated AI-positive ECGs; (3) Reviewing all user-initiated ECGs; (4) Reviewing all ECGs; (5) reviewing AI-negative ECGs.

Patient-level economic burden was assessed by comparing cumulative time and direct medical costs across AI-ECG mobile screening system, in-hospital 12-lead ECG, and electrophysiology study (EPS), highlighting differences in time \cite{R30, R31, R32, R33}and financial costs \cite{BeijingInsurance2021} under alternative screening pathways.

\subsection*{Statistical Analysis}
Model discrimination was evaluated using the area under the ROC curve (AUROC), with complementary assessment based on confusion matrices. Calibration was examined using calibration curves and quantified by the Brier score. Given the low prevalence of WPW in real-world screening, a positive-sample–focused score stability analysis was performed. Based on the Youden index, individuals in the highest 15\% of predicted risk (85th percentile) were defined as high risk, and the mean predicted probability among true-positive cases within this stratum was used to assess score concentration and internal consistency.

The clinical value of AI alerts was quantified using relative risk (RR). Workload reduction was assessed by the proportion of ECGs not requiring cardiologist review under the AI-assisted workflow. Screening efficiency across different strategies was evaluated using the NNR, defined as the inverse of the PPV.

In baseline comparisons, categorical variables are presented as counts and percentages, and continuous variables as means with standard deviations. Student’s t test or the Mann–Whitney U test was applied as appropriate. All analyses were performed using Python (version 3.9) with matplotlib, with a two-sided P value < 0.05 considered statistically significant.

\section*{Results}
\subsection*{Study Population}
Between August 2019 and May 2025, 4,184,097 raw single-lead ECG recordings were collected nationwide in China. After excluding 617,471 low-quality recordings, 3,566,626 ECGs from 87,836 individuals were included in the final analysis.

At the first stratification level, 16,457 ECGs (0.46\%) were classified as AI-positive for WPW , while 3,550,169 (99.54\%) were classified as non-WPW. At the second level, records were stratified based on whether users initiated cardiologist review. Among 16,457 AI-positive ECGs, 1,984 (12.05\%) underwent user-initiated cardiologist review, compared with 46,120 of 3,550,169 AI-negative ECGs (1.30\%), resulting in 48,104 reviewed ECGs (1.35\% of the total). Third-level analysis focused on the reviewed AI-negative ECGs, among which 9,750 (21.14\%) were initially classified as sinus normal and 36,370 (78.86\%) as other cardiac conditions.

For baseline comparisons (Table 1), two cohorts were defined based on cardiologist review status (review group, N = 15,483; non-review group, N = 68,759). The two groups were  comparable with respect to age, sex, height, blood pressure, and other physiological characteristics, with no clinically meaningful differences observed.  However the review group exhibited a substantially higher comorbidity burden, including hypertension (6.9\% vs 0.9\%), coronary stenting (2.4\% vs 0.2\%), myocardial infarction (0.6\% vs 0.1\%), and diabetes (2.5\% vs 0.4\%).In terms of lifestyle factors, the review group also demonstrated substantially higher participation in physical activity (28.9\% vs 2.0\%). These findings indicate that user-initiated cardiologist review is associated with greater underlying health risk and health awareness.

\renewcommand{\arraystretch}{1.2} 
\small 

\begin{longtable}{lccc}
\caption{Baseline characteristics of the study population stratified by cardiologist review status. Data are presented as arithmetic mean (SD) and n (\%).} \label{tab:baseline_characteristics} \\

\toprule
\textbf{Variable} & \begin{tabular}[c]{@{}c@{}}\textbf{Non-cardiologist}\\\textbf{review group}\end{tabular} & \begin{tabular}[c]{@{}c@{}}\textbf{Cardiologist}\\\textbf{review group}\end{tabular} & \textbf{$P$ value} \\ 

\midrule
\endfirsthead


\toprule
\textbf{Variable} & \begin{tabular}[c]{@{}c@{}}\textbf{Non-cardiologist}\\\textbf{review group}\end{tabular} & \begin{tabular}[c]{@{}c@{}}\textbf{Cardiologist}\\\textbf{review group}\end{tabular} & \textbf{$P$ value} \\ 
\midrule
\endhead

\midrule
\endfoot

\bottomrule
\endlastfoot

Age (year) & 49 (17) & 50 (16) & 0.7096 \\
\addlinespace
Gender (\%) & & & \\
\quad Male & 11,236 (16.5\%) & 4,299 (27.8\%) & \multirow{3}{*}{0.3386} \\
\quad Female & 7,355 (10.8\%) & 2,738 (17.7\%) & \\
\quad No report & 50,168 (72.7\%) & 8,446 (54.5\%) & \\
\addlinespace
Height (cm) & 167.7 (9.7) & 168.8 (9.5) & 0.6026 \\
Weight (kg) & 68.7 (14.9) & 68.5 (13.9) & $ < 0.01 $ \\
Systolic pressure (mmHg) & 117.5 (21.9) & 119.2 (21.1) & 0.3703 \\
Diastolic pressure (mmHg) & 85.2 (20.8) & 85.6 (19.9) & 0.8081 \\
\addlinespace
Hypertension (\%) & & & \\
\quad Yes & 611 (0.9\%) & 1,065 (6.9\%) & \multirow{3}{*}{$ < 0.0001 $} \\
\quad No & 3,830 (5.6\%) & 4,391 (28.4\%) & \\
\quad No report & 64,318 (93.5\%) & 10,027 (64.8\%) & \\
\addlinespace
Congenital heart disease (\%) & & & \\
\quad Yes & 45 (0.1\%) & 80 (0.5\%) & \multirow{3}{*}{$ < 0.05 $} \\
\quad No & 4,386 (6.4\%) & 5,373 (34.7\%) & \\
\quad No report & 64,328 (93.5\%) & 10,030 (64.8\%) & \\
\addlinespace
Heart stent (\%) & & & \\
\quad Yes & 123 (0.2\%) & 225 (1.5\%) & \multirow{3}{*}{$ < 0.001 $} \\
\quad No & 4,309 (6.3\%) & 5,228 (33.7\%) & \\
\quad No report & 64,327 (93.5\%) & 10,030 (64.8\%) & \\
\addlinespace
Atrial fibrillation (\%) & & & \\
\quad Yes & 349 (0.5\%) & 20 (0.1\%) & \multirow{3}{*}{0.2651} \\
\quad No & 4,081 (5.9\%) & 5,433 (35.1\%) & \\
\quad No report & 64,329 (93.6\%) & 10,030 (64.8\%) & \\
\addlinespace
Myocardial infarction (\%) & & & \\
\quad Yes & 51 (0.1\%) & 98 (0.6\%) & \multirow{3}{*}{$ < 0.01 $} \\
\quad No & 4,379 (6.4\%) & 5,355 (34.6\%) & \\
\quad No report & 64,329 (93.5\%) & 10,030 (67.8\%) & \\
\addlinespace
Diabetes (\%) & & & \\
\quad Yes & 248 (0.4\%) & 389 (2.5\%) & \multirow{3}{*}{$ < 0.01 $} \\
\quad No & 4,189 (6.1\%) & 5,065 (32.7\%) & \\
\quad No report & 64,322 (93.5\%) & 10,029 (67.8\%) & \\
\addlinespace
Kidney disease (\%) & & & \\
\quad Yes & 49 (0.1\%) & 74 (0.5\%) & \multirow{3}{*}{0.2582} \\
\quad No & 4,381 (6.4\%) & 5,379 (34.7\%) & \\
\quad No report & 64,329 (93.5\%) & 10,030 (67.8\%) & \\
\addlinespace
Pacemaker (\%) & & & \\
\quad Yes & 46 (0.1\%) & 24 (0.1\%) & \multirow{3}{*}{$ < 0.001 $} \\
\quad No & 4,387 (6.4\%) & 5,429 (35.1\%) & \\
\quad No report & 64,326 (93.5\%) & 10,030 (67.8\%) & \\
\addlinespace
Habit (\%) & & & \\
\quad Drink & 621 (0.9\%) & 1,507 (9.7\%) & 0.0872 \\
\quad Smoke & 261 (0.4\%) & 1,807 (11.7\%) & 0.3362 \\
\quad Drinking tea/coffee & 759 (1.1\%) & 1,578 (10.2\%) & $ < 0.05 $ \\
\quad Sport & 1,400 (2.0\%) & 4,289 (27.7\%) & $ < 0.001 $ \\

\end{longtable}

\subsection*{AI Diagnostic Performance}
The model demonstrated moderate discriminative performance for WPW detection, with an AUC of 0.6676 (Figure 3A), which was statistically significantly better than random classification. Within the cardiologist-reviewed cohort, among 1,984 ECGs initially classified as WPW by the AI model, 25 were confirmed as true positives (TP) and 1,959 as false positives (FP). Conversely, among 46,120 ECGs classified as non-WPW, 30 were false negatives (FN) and 46,090 were true negatives (TN),as shown in Figure 3B. Based on these results, the overall sensitivity and specificity of the model were 45.5\% and 95.92\%, respectively.

The impact of comorbid cardiac conditions on model performance was further assessed. Among 55 confirmed WPW cases, 5 had concomitant cardiac conditions, of which only 1 was correctly identified by the AI model, resulting in a sensitivity of 20\%. In contrast, 24 of 50 WPW cases without comorbidities were correctly detected, corresponding to a higher sensitivity of 48\%. As for specificity, among 48,049 true non-WPW cases, 29,644 presented with other cardiac conditions, of which 29,385 were correctly classified as non-WPW (specificity, 99.13\%). By comparison, among 18,405 records without comorbidities, 16,705 were correctly classified, resulting in a lower specificity of 90.76\%.

\begin{table}[htbp]
\centering
\begin{threeparttable}
\caption{Detailed diagnostic performance of the AI model stratified by comorbidities.}
\label{tab:ai_performance}
\small 
\begin{tabularx}{\textwidth}{l r r r r r r r} 
\toprule
\textbf{Subgroup} & \textbf{Total N} & \textbf{TP} & \textbf{FN} & \textbf{TN} & \textbf{FP} & \textbf{Sens (\%)} & \textbf{Spec (\%)} \\ 
\midrule
Overall Performance & 48,104 & 25 & 30 & 46,090 & 1,959 & 45.50 & 95.92 \\ 
\midrule
\multicolumn{8}{l}{\textit{WPW Cases}} \\
\hspace{1em}Without Comorbidities & 50 & 24 & 26 & -- & -- & 48.00 & -- \\
\hspace{1em}With Comorbidities\tnote    & 5 & 1 & 4 & -- & -- & 20.00 & -- \\ 
\midrule
\multicolumn{8}{l}{\textit{Non-WPW Cases}} \\
\hspace{1em}Without Comorbidities & 18,405 & -- & -- & 16,705 & 1,700 & -- & 90.76 \\
\hspace{1em}With Comorbidities\tnote & 29,644 & -- & -- & 29,385 & 259 & -- & 99.13 \\ 
\bottomrule
\end{tabularx}

\begin{tablenotes}
    \footnotesize
    \item Abbreviations: TP, True Positive; FN, False Negative; TN, True Negative; FP, False Positive; WPW, Wolff-Parkinson-White syndrome.
\end{tablenotes}
\end{threeparttable}
\end{table}

Regarding calibration performance(Figure 3C), the Brier score was 0.1694, with a calibration slope of 0.1989 and an intercept of -6.6583, indicating substantial global miscalibration and probability overestimation under conditions of extremely low disease prevalence. However, true-positive cases within the high-risk stratum (top 15\% of predicted risk) exhibited a high mean predicted score (0.9154),as shown in Figure 3D, indicating strong internal consistency in identifying WPW despite biased absolute probability estimates.

\begin{figure}[htbp] 
    \centering
    \includegraphics[width=0.8\textwidth]{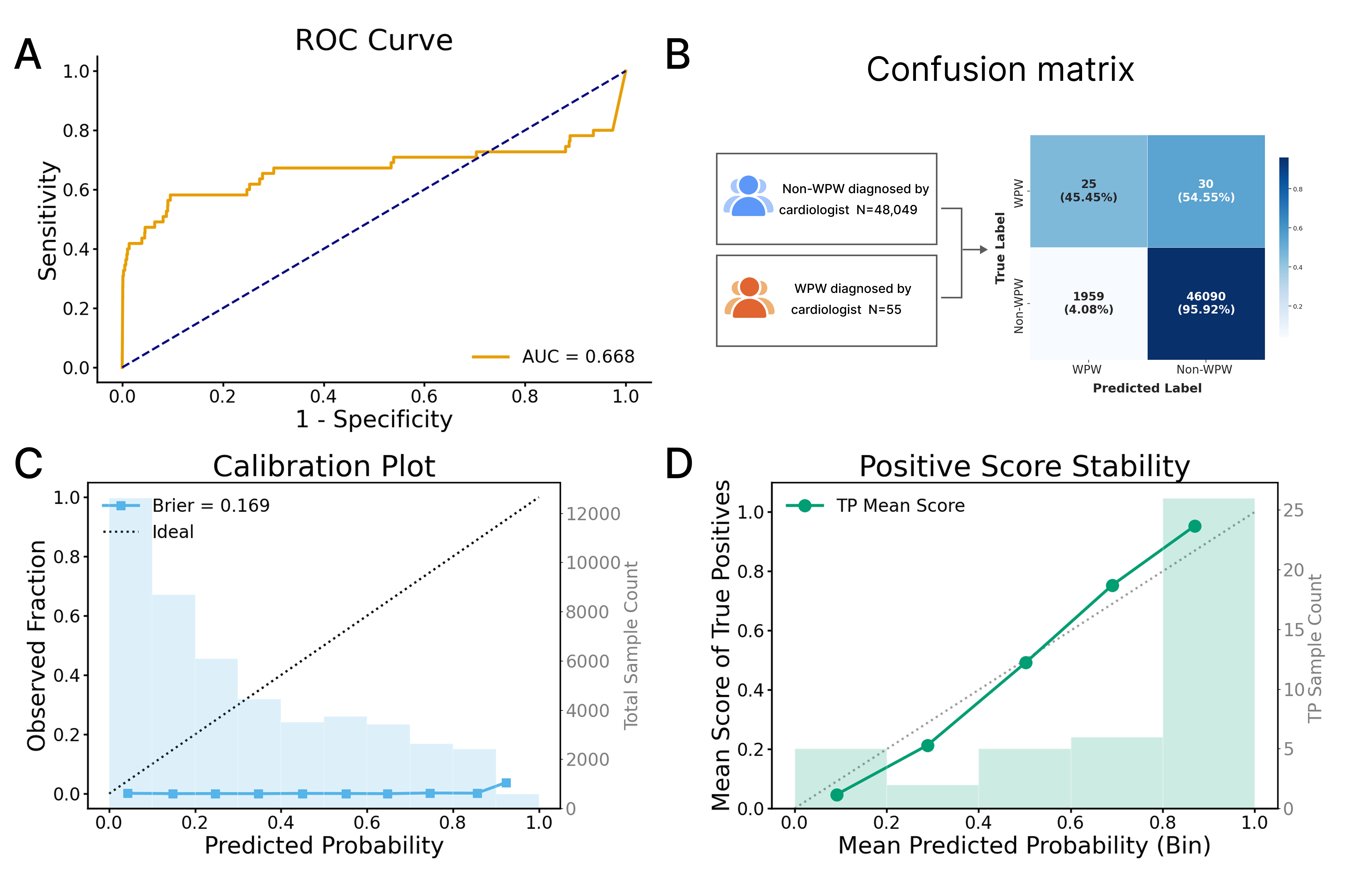} 
    \caption{Diagnostic performance and risk stratification capability of the AI model in real-world settings. (A) Receiver Operating Characteristic (ROC) curve illustrating the overall discriminative performance of the AI model for WPW detection. (B) Confusion matrix  showing the distribution of True Positives, False Positives, False Negatives, and True Negatives. (C) Calibration curve assessing the agreement between predicted probabilities and observed frequencies, highlighting the global probability overestimation caused by the extremely low disease prevalence. (D) Score stability analysis for confirmed True Positive cases, demonstrating the concentration of predictive confidence scores within the high-risk stratum.}
    \label{AI}
\end{figure}

\subsection*{User Behavior and Cardiologist Review}
User behavior analysis focused on the decision to seek cardiologist review after receiving AI-generated diagnoses (Figure 4). Among a total of 3,566,626 records, 48,104 (1.35\%) underwent user-initiated cardiologist review. This decision was strongly influenced by the initial AI output. Of the 16,457 records classified as WPW by the AI, 1,984 underwent review, resulting in a review rate of 12.05\%.In contrast, only 46,120 of 3,550,169 AI-negative records underwent review, corresponding to a much lower rate of 1.30\%. This indicates that users who received an AI-positive warning were nearly nine times more likely to seek cardiologist confirmation compared to others.

Analysis of the motivations behind the 46,120 review requests in the AI-negative group revealed that the primary driver was the detection of other AI-identified abnormalities. Data showed that the vast majority of these requests (78.86\%; n = 36,370) were for records diagnosed by the AI as other cardiac conditions. This is consistent with the previous baseline cohort analysis, suggesting that underlying health risks increase user review-seeking behavior. Simultaneously, a significant proportion of users (21.14\%; n = 9,750) still sought a cardiologist review even when the AI reported sinus normal.

Regarding AI-positive WPW records, an analysis of the 1,984 records reviewed by cardiologists identified 25 true WPW cases, resulting in a Positive Predictive Value (PPV) of 1.26\%. Conversely, among the 46,120 reviewed records that the AI-negative for WPW, only 30 were eventually confirmed as WPW—a significantly lower rate of 0.065\%. This disparity highlights the overall reliability of the AI across different diagnostic categories.

To further assess potential user-selection bias, 500 records were randomly sampled from 14,473 AI-positive records that did not undergo user-initiated review. The cardiologist review confirmed 28 of these as true WPW cases, resulting in a confirmation rate of 5.6\%.

\begin{figure}[htbp] 
    \centering
    \includegraphics[width=0.8\textwidth]{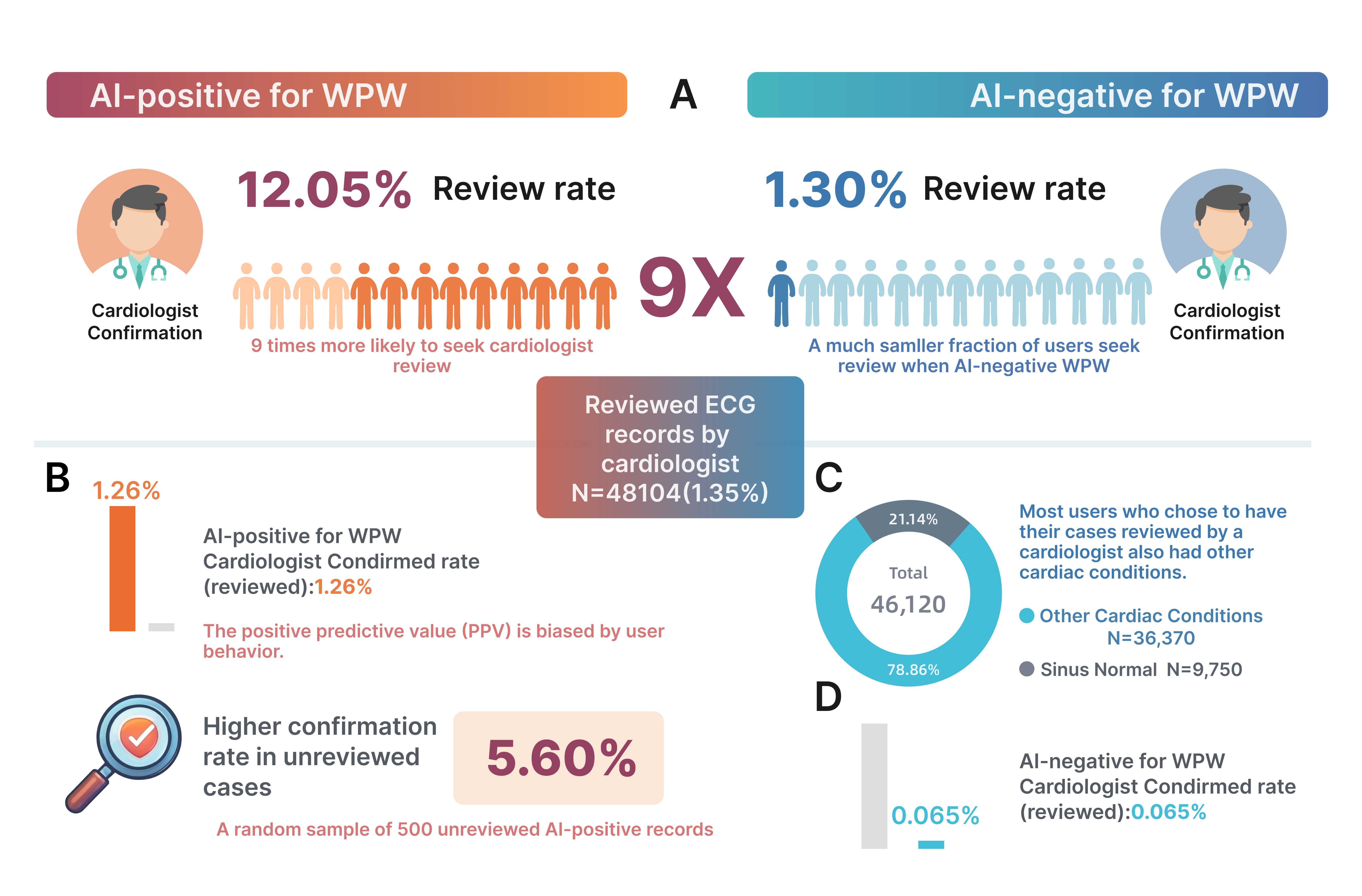} 
    \caption{Dynamics of user behavior and selection bias in the human-AI collaborative workflow. (A) Comparison of cardiologist review rates stratified by AI results, demonstrating that AI-positive alerts serve as a strong behavioral driver for seeking medical advice. (B) Comparison of the Positive Predictive Value (PPV) in the user-initiated AI-positive cohort and the confirmation rate in the randomly sampled non-reviewed cohort. (C) Composition of diagnoses within the AI-negative review cohort, revealing that the detection of "other cardiac conditions" is the primary motivator for user-initiated reviews. (D) Confirmation rate in the AI-negative cohort, validating the model's robust negative predictive capability and low risk of missed detection.}
    \label{HCI}
\end{figure}

\subsection*{Value of Human-AI Collaboration}
The efficiency gains achieved through AI-assisted screening are of substantial practical significance. Relative risk (RR) analysis showed that the detection rate of WPW was 5.6\% among ECGs classified as AI-positive, compared with only 0.065\% among AI-negative records, corresponding to a remarkably high relative risk of 86.2. This indicates that the AI model effectively enriched high-risk individuals.

Under the baseline scenario without AI assistance, cardiologists would be required to manually review all 3,566,626 ECG records. In contrast, under the AI-assisted model, cardiologists only need to prioritize the evaluation of high-risk samples filtered by the AI. Even if all 16,457 AI-positive records were reviewed, the total review workload would still be reduced by approximately 99.5\% compared with the baseline.

These findings demonstrate that the collaborative model combining AI screening with cardiologist review not only substantially reduces manual review burden but also realizes effective risk stratification and enhanced screening efficiency within large-scale opportunistic population screening.

\subsection*{System Efficiency and Effectiveness}
To evaluate the effectiveness of the screening system, the total number of WPW cases in the dataset was estimated. Based on the 5.6\% PPV derived from random sampling, approximately 922 true WPW cases were inferred among the 16,457 AI-positive records. Meanwhile, the 0.065\% missed detection rate observed in the reviewed AI-negative cohort extrapolated to the entire 3,550,169 AI-negative records suggests roughly 2,308 missed cases. Together, these estimates indicate approximately 3,230 WPW cases in the full dataset.

As illustrated in Figure 4A, based on these estimations, the baseline screening effectiveness for a cardiologist reviewing the entire population is approximately one confirmed case per 1,104 screenings (NNR = 1,104). If cardiologists only review the 48,104 records that underwent user-initiated review, one WPW case is confirmed for every 875 records. Furthermore, if review efforts are concentrated solely on user-initiated AI-positive records, the NNR decreases significantly to 79. This approximately 14-fold difference in workload underscores the efficiency gains provided by the integrated opportunistic screening system.In an ideal scenario where all 16,457 AI-positive WPW records are reviewed, cardiologists would need only 18 records to confirm a true WPW case, corresponding to a 61-fold increase in efficiency. Conversely, reviewing all 3,550,169 AI-negative records is highly inefficient, requiring an average of 1,538 records per detected missed case.
\begin{table}[htbp]
    \centering
    \caption{Summary of Overall Efficiency and Effectiveness Results.}
    \label{tab:efficiency_effectiveness}
    \small 
    \renewcommand{\arraystretch}{1.5} 
    \begin{tabularx}{\textwidth}{p{3.2cm} l l >{\raggedright\arraybackslash}X}
        \toprule
        Assessment Category & Metric & Value & Notes and Calculation Basis \\ 
        \midrule
        
        Efficiency Gain & Expert Workload Reduction & $\sim$99.5\% & Based on AI-positive rate of 0.46\% (1 - 16,457 / 3,566,626). \\
        
        \midrule
        
        \multirow{3}{3.2cm}{Prevalence and Case Estimation} 
          & Estimated AI-Detected & $\sim$922 cases & 16,457 (AI-positive) $\times$ 5.6\% (Positive Predictive Value) \\
          & Estimated Missed Cases & $\sim$2,308 cases & 3,550,169 (AI-negative) $\times$ 0.065\% \\
          & Total Estimated Cases & $\sim$3,230 cases & Sum of detected and missed; Estimated Prevalence $\sim$0.9 per 1,000. \\
        
        \midrule
        
        \multirow{5}{3.2cm}{Screening Efficiency (NNR)} 
          & Strategy 1: Baseline & $\sim$1,104 & Universal manual review. \\
          & Strategy 2: User-Initiated & $\sim$875 & Review all user-initiated (48,104 reviews / 55 cases). \\
          & Strategy 3: AI-Collaborative & $\sim$79 & Review user-initiated AI-positives (1,984 reviews / 25 cases). \\
          & Strategy 4: AI-Ideal & $\sim$18 & Review all AI-positives (1 / 5.6\% Positive Predictive Value). \\
          & Strategy 5: AI-Negative & $\sim$1,538 & Review AI-negatives (1 / 0.065\%). \\
        
        \bottomrule
    \end{tabularx}
    
    \footnotesize
    \vspace{1ex}
    \textit{Note: NNR = Number Needed to Review to confirm one case. }
\end{table}

From the user perspective, the AI-ECG mobile screening system offers substantial time and cost advantages. For example, in Beijing, a single hospital-based 12-lead ECG typically costs 80 RMB or more, while EPS range from approximately 1,200 to 2,500 RMB. By comparison, the initial hardware cost for the AI-ECG system is approximately 100 RMB, and subsequent online consultations with a cardiologist cost about 9 RMB, markedly reducing cumulative expenditures for long-term monitoring (Figure 5B). In terms of time, traditional hospital-based ECG testing (including scheduling, consultation, and examination) requires roughly 140 minutes, whereas a single AI-ECG mobile screening can be completed in approximately 10 minutes, representing a 92.9\% reduction in time (Figure 5C).

\begin{figure}[htbp] 
    \centering
    \includegraphics[width=0.8\textwidth]{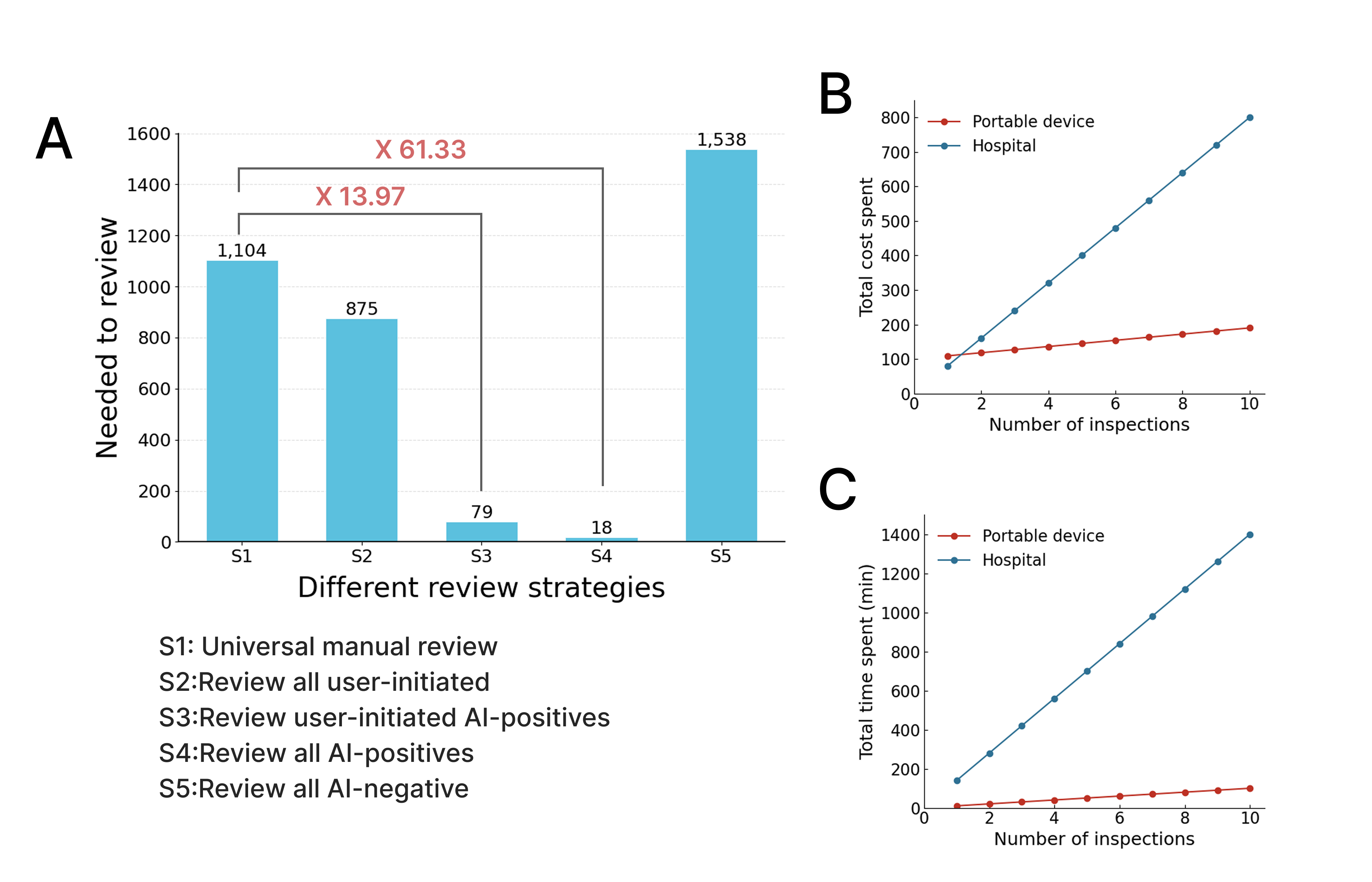} 
    \caption{Comparative analysis of screening efficiency and cumulative resource burdens. (A) Comparison of screening efficiency measured by Number Needed to Review (NNR) across different strategies. The human-AI collaborative workflow significantly lowered the NNR compared to universal manual screening. (B) Cumulative cost comparison between the AI-ECG mobile system and traditional hospital procedures over 10 sequential screening sessions, highlighting the widening cost-saving trend. (C) Cumulative time consumption comparison between the AI-ECG mobile system and traditional hospital procedures over 10 sequential screening sessions.}
    \label{NNR}
\end{figure}

This favorable cost-effectiveness makes the AI-ECG mobile screening system an economically feasible tool for large-scale opportunistic WPW screening. We illustrate this with a representative case in which a 42-year-old asymptomatic male performed a routine home ECG (Figure 6). The AI model identified typical pre-excitation features within XX seconds of data upload and issued a high-risk alert. The user subsequently initiated a consultation within the application and received a confirmed diagnosis. This case highlights the capability of the AI-ECG mobile screening system to detect WPW in asymptomatic individuals.

\begin{figure}[htbp] 
    \centering
    \includegraphics[width=0.8\textwidth]{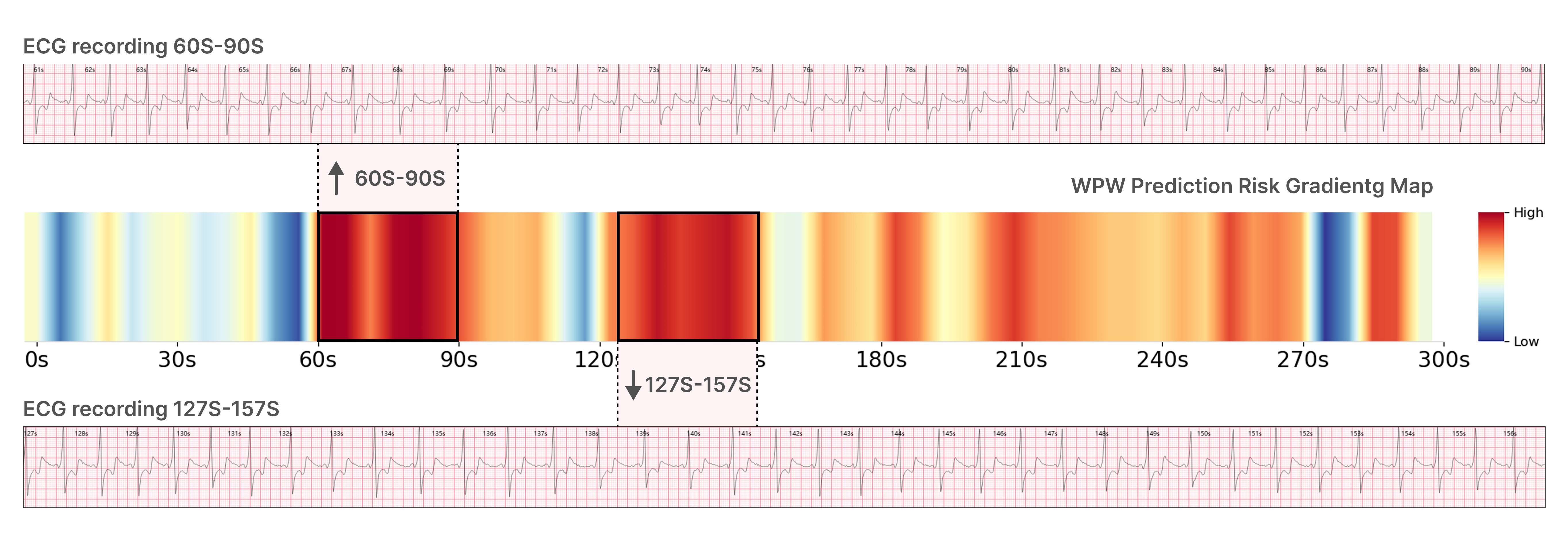} 
    \caption{ Representative case: AI-driven detection of WPW syndrome in an asymptomatic individual.}
    \label{case}
\end{figure}

\section*{Discussion}
This study systematically evaluated the effectiveness of a closed-loop screening paradigm—\\portable device, AI-based primary screening, and cardiologist review—for opportunistic detection of WPW syndrome using large-scale real-world data. The main findings indicate that this paradigm is not only technically feasible, effectively enriching high-risk individuals through AI, but also significantly reduces healthcare resource utilization and demonstrates substantial public health value in terms of cost and time efficiency.

\subsection*{AI Diagnostic Performance and Risk Stratification}
The extremely low prevalence of WPW in the general population (approximately 1–3 per 1,000) makes traditional large-scale population screening cost-inefficient. In this study, the AI model demonstrated moderate overall performance (AUC 0.6676) under complex real-world signal conditions, while achieving very high specificity (95.92\%). Calibration analysis indicated a global tendency to overestimate risk due to class imbalance; however, among the top 15\% high-risk stratum, the mean predicted score for true positives reached 0.9154. These results suggest that, despite modest traditional performance metrics, the clinical utility of the model lies primarily in its ability to stratify risk and enrich potential high-risk individuals from a large low-risk population, providing precise targets for cardiologist review.

Notably, overall sensitivity was 45.5\%, but higher (48\%) in WPW cases without comorbidities and markedly lower (20\%) in cases with other cardiac conditions. Although the latter subgroup was small (n = 5) and should be interpreted cautiously, this phenomenon indicates that complex pathophysiological backgrounds may obscure typical pre-excitation features, such as Delta waves, complicating AI detection\cite{R35}. These findings further underscore the necessity of human–AI collaborative workflows in real-world clinical settings.

\subsection*{Efficiency gains of human–AI collaboration}
The most significant finding of this study is the substantial improvement in screening efficiency achieved through the AI–physician collaborative model. Under manual screening of the entire population, an average of 1,104 ECG records must be reviewed to confirm a single WPW case. By contrast, in an ideal scenario where all AI-positive records undergo cardiologist review, only approximately 18 records need to be examined to confirm one case, representing a 61-fold efficiency gain. Relative risk analysis further confirmed that the confirmation rate among AI-positive records was 86.2 times higher than in AI-negative records. By implementing this risk-stratified collaborative workflow, the cardiologist review workload is reduced by about 99.5\%, allowing limited expert resources to focus on samples with the highest likelihood of true positive findings.

\subsection*{User behavior and selection bias}
In real-world settings, user behavior significantly influences screening outcomes. Users receiving an AI-positive alert were nine times more likely to seek cardiologist review than those with AI-negative results (12.05\% vs. 1.30\%), indicating that AI output strongly drives the decision to request cardiologist evaluation. Baseline comparisons further revealed that users who proactively sought review exhibited a higher burden of comorbidities (e.g., hypertension, history of coronary stenting) and more active lifestyles (higher exercise participation), suggesting that prior health status and risk perception collectively shape user behavior.

However, this self-selection introduces bias. Specifically, the PPV among voluntarily reviewed AI-positive cases was only 1.26\%, markedly lower than the 5.6\% observed in the randomly sampled review cohort. This discrepancy likely reflects that users requesting review are more often driven by unclear ECG features, subjective symptom perception, or the presence of other cardiac conditions, diluting the proportion of true positives in the review sample. These findings provide insight for future strategies to optimize screening workflows through user behavior–driven design.

\subsection*{Limitations}
This study has several limitations. First, it was based on single-lead ECG data, which may lack information on certain vector-specific accessory pathway features compared with standard 12-lead ECGs, potentially contributing to the observed sensitivity limitations. Second, user-initiated review introduces selection bias. Although we applied random sampling and multi-layer stratified analyses to mitigate this, there may still be unquantified missed cases within the AI-negative cohort that did not undergo review. Finally, as a retrospective study, despite the large dataset, future prospective clinical trials are still required to further validate the long-term benefits of this screening model in reducing the incidence of SCD.

\subsection*{Future Outlook}
Building on these findings, future research could explore several avenues. Large-scale prospective cohort studies are needed to longitudinally follow screened WPW patients and assess the real-world impact of this screening model on hard clinical outcomes. Additionally, future work could investigate how AI alerts influence individual health behaviors and cardiologist decision-making, providing deeper insights into the interactions among AI, users, and clinicians.

This study introduces a cost-effective paradigm for opportunistic screening with implications beyond early WPW detection. Portable ECG devices, characterized by low cost, high accessibility, and the ability to capture paroxysmal arrhythmias in daily life, possess a natural advantage as gateways for long-term health monitoring. Supported by internet and telemedicine technologies, this paradigm integrates physiological data acquisition, AI analysis, user feedback, and expert intervention. It outlines a new form of telemedicine service centered on continuous monitoring and risk stratification.

\section*{Acknowledgments}
This work is supported by the National Natural Science Foundation of China (62102008, 62172018), CCF-Tencent Rhino-Bird Open Research Fund (CCF-Tencent RAGR20250108), CCF-Zhipu Large Model Innovation Fund (CCF-Zhipu202414), PKU-OPPO Fund (BO202301, BO202503), Research Project of Peking University in the State Key Laboratory of Vascular Homeostasis and Remodeling (2025-SKLVHR-YCTS-02). Fan Sumei scientific research start-up funds (grant No. DT2400000509)

\section*{Competing interests}
The authors declare no competing interests.

\newpage

\bibliography{reference_cleaned}

\end{document}